\documentclass{article}

\usepackage{arxiv}

\usepackage[utf8]{inputenc} 
\usepackage[T1]{fontenc}    
\usepackage{natbib}
\usepackage{hyperref}       
\usepackage{url}            
\usepackage{booktabs}       
\usepackage{amsfonts}       
\usepackage{nicefrac}       
\usepackage{microtype}      
\usepackage{lipsum}		
\usepackage{graphicx}
\usepackage{doi}
\usepackage{amsmath}

\title{Black hole triggered star formation in the dwarf galaxy Henize 2-10}

\author{ \href{https://orcid.org/0000-0001-7412-8988}{\includegraphics[scale=0.06]{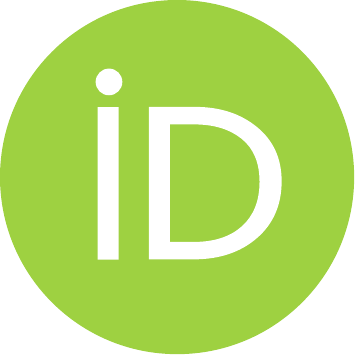}\hspace{1mm}Zachary Schutte} \\
	eXtreme Gravity Institute, Department of Physics\\
	Montana State University\\
	Bozeman, MT 59715, USA \\
	\texttt{zachary.schutte@montana.edu} \\
	\And
	\href{https://orcid.org/0000-0001-7158-614X}{\includegraphics[scale=0.06]{orcid.pdf}\hspace{1mm}Amy Reines} \\
	eXtreme Gravity Institute, Department of Physics\\
	Montana State University\\
	Bozeman, MT 59715, USA \\
	\texttt{amy.reines@montana.edu} \\
}

\renewcommand{\shorttitle}{Black hole triggered star formation in the dwarf galaxy Henize 2-10}

\hypersetup{
pdftitle={Black hole triggered star formation in the dwarf galaxy Henize 2-10},
pdfauthor={Zachary Schutte, Amy Reines},
}

\begin{document}
\maketitle

\textbf{Black hole driven outflows have been observed in some dwarf galaxies with active galactic nuclei \cite{manzano2019agn}, and likely play a role in heating and expelling gas (thereby suppressing star formation), as they do in larger galaxies \cite{fabian2012observational}. The extent to which black hole outflows can trigger star formation in dwarf galaxies is unclear, because work in this area has hitherto focused on massive galaxies and the observational evidence is scarce \cite{gaibler2012jet,maiolino2017star,gallagher2019widespread}. Henize 2-10 is a dwarf starburst galaxy previously reported to have a central massive black hole\cite{reines2011actively_Henize,reines2012parsec,reines2016deep,riffel2020evidence} though that interpretation has been disputed since some aspects of the observational evidence are also consistent with a supernova remnant \cite{hebbar2019x,cresci2017muse}. At a distance of $\sim$9 Mpc, it presents an opportunity to resolve the central region and determine if there is evidence for a black hole outflow impacting star formation. Here we report optical observations of Henize 2-10 with a linear resolution of a few parsecs. We find a $\sim$150 pc long ionized filament connecting the region of the black hole with a site of recent star formation.  Spectroscopy reveals a sinusoid-like position-velocity structure that is well described by a simple precessing bipolar outflow.  We conclude that this black hole outflow triggered the star formation.} \par

Radio observations of Henize 2-10 using very long baseline interferometry reveal a nuclear, compact, non-thermal source with a luminosity of L$_R \sim 4 \times 10^{35}$ erg s$^{-1}$ and a physical size $<$ 3 pc × 1 pc \cite{reines2012parsec}. High-resolution X-ray observations unveil a point source with L$_X \sim 10^{38}$ erg s$^{-1}$ that is spatially coincident with the compact nuclear radio source \cite{reines2016deep}. There are two possible explanations for these radio and X-ray observations alone - a highly sub-Eddington massive black hole (i.e., a low-luminosity AGN) or a very young supernova remnant\cite{reines2012parsec,hebbar2019x}. However, there are other observational results to consider regarding the origin of the nuclear radio/X-ray source in Henize 2-10. We summarize these results in Figure \ref{tab:AGNSNR_tab} (Extended Data Table 1) and demonstrate that a highly sub-Eddington massive black hole is consistent with all the available observations including new results presented here, while a supernova remnant is not. \par

\begin{figure}
\centering
\includegraphics[width=0.85\textwidth]{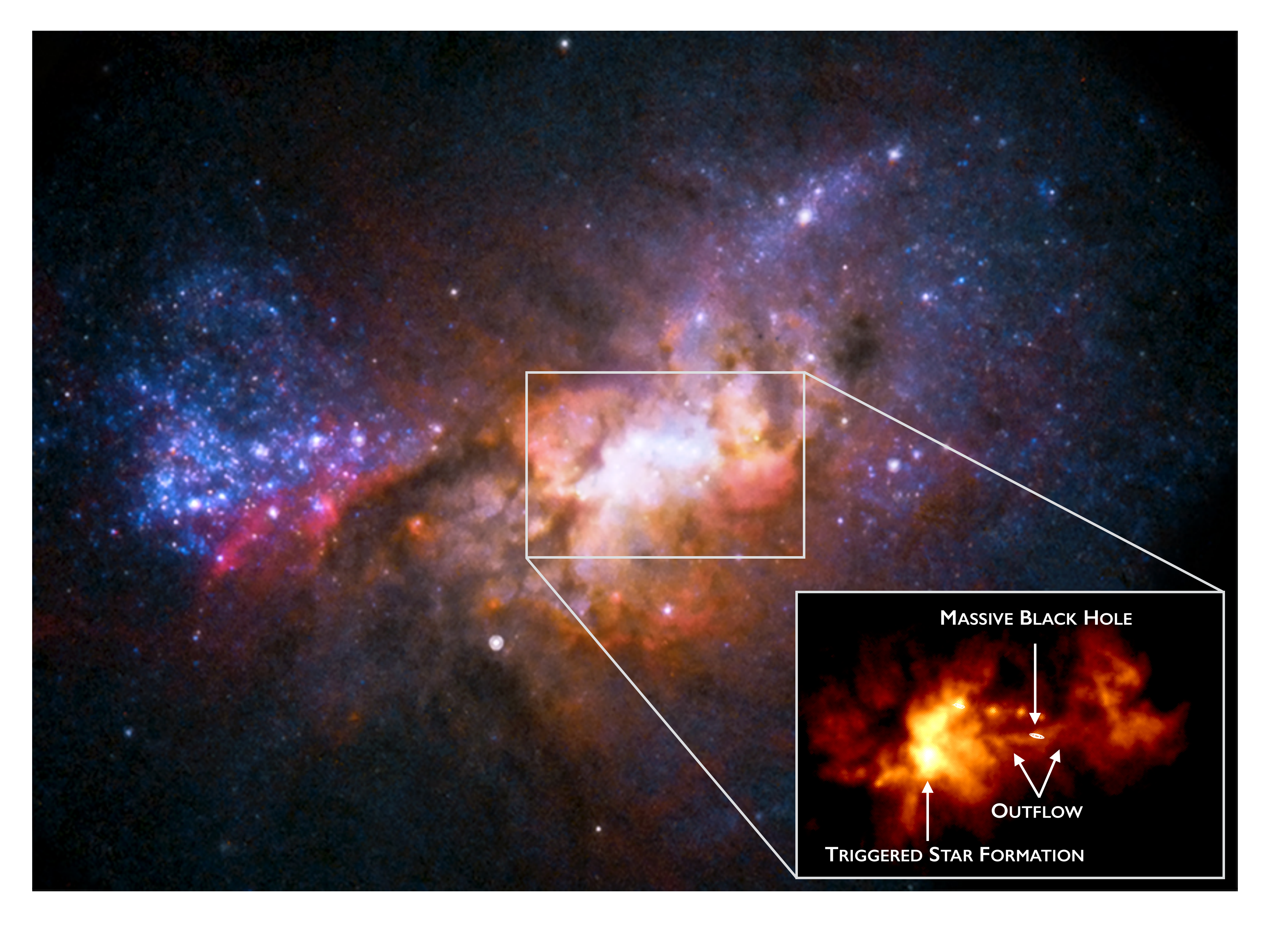}
\caption{\textbf{HST optical image of the dwarf starburst galaxy Henize 2-10 (Image credit: NASA/STScI).} The inset shows a narrowband H$\alpha+$continuum image of the central 6” x 4” region. At the distance of Henize 2-10 ($\sim$9 Mpc), 1” corresponds to $\sim$ 44 pc. White contours indicate compact radio emission detected with very long baseline interferometry and mark the location of the central massive black hole \cite{reines2012parsec}, which is also detected in X-rays \cite{reines2016deep}. Our HST spectroscopy demonstrates that the black hole is driving an outflow that is triggering the formation of young massive star clusters. The main image is 25” ($\sim$1.1 kpc) across.}
\label{fig:H210_photo}
\end{figure}

We observed Henize 2-10 at optical wavelengths using the Space Telescope Imaging Spectrograph (STIS) on the Hubble Space Telescope (HST). We obtained observations of the central regions of Henize 2-10 with the 0.2"-slit in two orientations. The first, referred to as the EW orientation, is centered on the nuclear radio/X-ray source and aligned with the filamentary ionized structure between the two bright, extended regions of ionized gas previously identified in narrowband H$\alpha$ imaging with HST \cite{reines2011actively_Henize} (Figure \ref{fig:H210_photo}). The second, referred to as the NS orientation, is centered on the nuclear source and rotated 90$^{\circ}$ with respect to the EW observation. We obtained high-dispersion observations (velocity resolution of $\sim 40$ km s$^{-1}$) at both slit positions using the G750M and G430M gratings, which cover the strong emission lines of interest (e.g., H$\alpha$, [NII], [SII], [OI] and H$\beta$, [OIII], respectively). \par


\begin{figure}
\centering
\includegraphics[width=\textwidth]{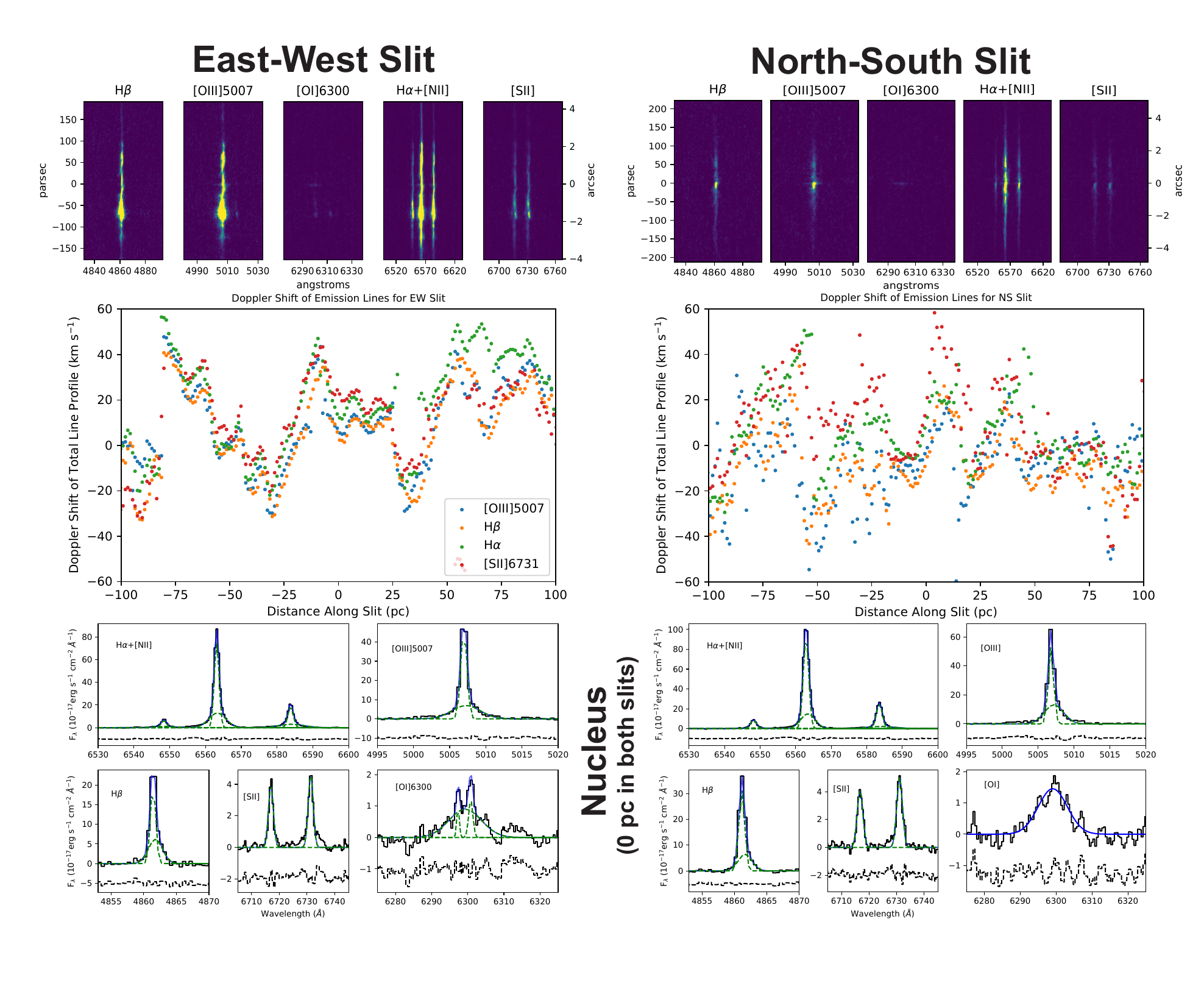}
\caption{\textbf{Optical spectra and ionized gas kinematics for the central region of Henize 2-10.} Top panels show 2D continuum-subtracted HST STIS spectra of key emission line regions along the EW and NS slit orientations. The middle panels show position-velocity diagrams along each slit position where Doppler shifts are relative to the systemic velocity of the galaxy (873 km s$^{-1}$) \cite{kobulnicky1995aperture}. The velocity pattern in the EW slit is much more coherent than that in the NS slit. The bottom panels show extracted 1D spectra at the location of the nuclear massive black hole. The region surrounding the nucleus (0 pc in both slit positions) shows strong broadened emission lines, including [OI]6300 and [OIII]5007.}
\label{fig:H210_AGN_spec1}
\end{figure}

The kinematics of the ionized gas provide evidence for an outflow originating from a nuclear massive black hole. First, we detect substantially broadened emission lines at the location of the central source in both slit orientations. In particular, the [OI]6300 emission line has a broad component with a full width at half maximum (FWHM) of 497 km s$^{-1}$ as measured in the EW slit (and 445 km s$^{-1}$ in the NS slit) using a 3-pixel extraction region in the spatial dimension.  While the [OI] line is too weak to be detected all along the EW filament (or NS slit), it is detected at the location of the bright star-forming region $\sim$70 pc to the east with a FWHM = 103 km s$^{-1}$, much less than that of the central source (e.g., see top left panel of Figure \ref{fig:H210_AGN_spec1}). The [OIII]5007 line is much stronger than [OI]6300 and detectable all along the EW filament.  The FWHM of the broad component of [OIII] at the location of the nuclear source is 271 km s$^{-1}$, which is somewhat broader than the median value per spatial pixel along the EW slit (175 km s$^{-1}$ with a standard deviation of 53 km s$^{-1}$). We do not observe such broad emission beyond the location of the central source in the NS slit (see top right panel of Figure \ref{fig:H210_AGN_spec1}).  We emphasize that the line widths at the location of the nuclear source are consistent with a low-velocity outflow from a massive black hole, but would be anomalously low for a very young supernova remnant with typical FWHMs of thousands of km s$^{-1}$ \cite{mathewson1980new,borkowski2017asymmetric}.  \par

In addition to broadened emission lines at the location of the central source, we also find Doppler shifted velocities along the EW slit that exhibit a coherent sinusoid-like pattern that is relatively smooth, especially in comparison to the position-velocity diagram along the NS slit orientation that shows no evidence of a coherent pattern in the Doppler velocities of strong emission lines (see Figure \ref{fig:H210_AGN_spec1} and Methods). Moreover, a simple model of a precessing bipolar outflow broadly reproduces the observed sinusoid-like velocity pattern along the EW slit that is aligned with the ionized filament seen in narrowband H$\alpha$ imaging with HST (see Methods for a description of the model). Precessing jets have been observed in many AGNs, although they are typically found in more luminous quasars and radio galaxies \cite{gower1982precessing,dunn2006precession}. Theories for the origin of precessing jets/outflows include accretion disk warping, jet instabilities, and the presence of massive black hole binaries \cite{pringle1996self,nixon2013jets}. On the other hand, the coherent velocity pattern we observe over $\sim$150 pc along the ionized filament centered on the central source is incompatible with a supernova remnant origin since supernova remnants do not drive quasi-linear outflows on such large scales. \par


\begin{figure}
\centering
\includegraphics[width=\textwidth]{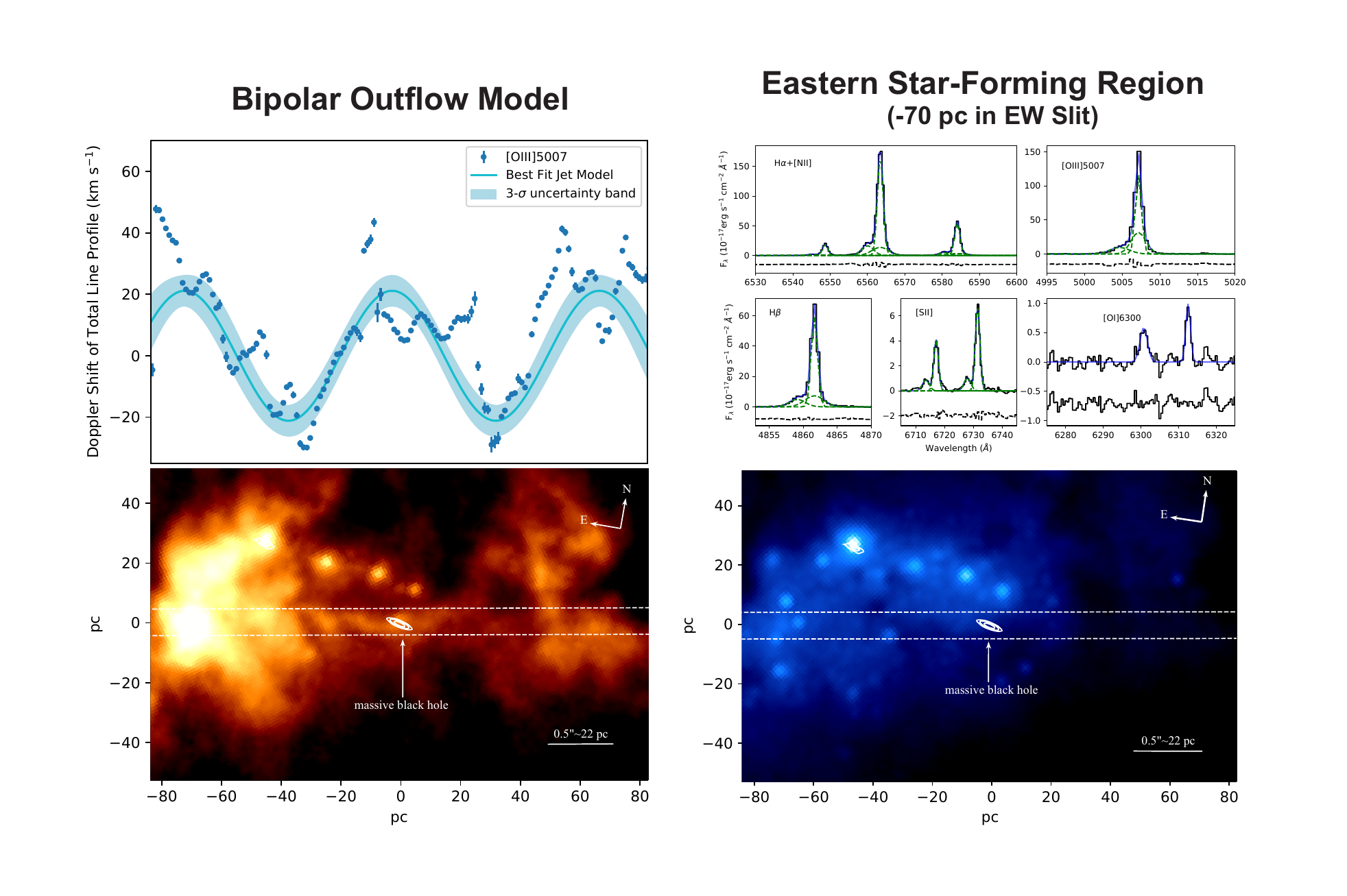}
\caption{\textbf{Visualization of the bipolar outflow model and star forming regions.}  Top left: Position-velocity diagram of the [OIII]5007 emission line along the EW slit where the Doppler shifts are relative to the systemic velocity of the galaxy (873 km s$^{-1}$) \cite{kobulnicky1995aperture}. The data are well described by a simple precessing outflow model with a precession frequency of f $\sim$ 5 revolutions Myr$^{-1}$ and a precession angle of $\theta$ $\sim$ 4$^\circ$ (see Methods). Bottom left: HST narrowband H$\alpha$+continuum image showing that the EW slit position is aligned with the ionized filamentary structure connecting the massive black hole and site of recent star formation ~70 pc to the east. White contours indicate compact radio emission from very long baseline interferometry \cite{reines2012parsec}, and show the location of the massive black hole. Bottom right: HST 0.8 micron ($\sim$I-band) broadband image with the same field of view. Young star clusters in the eastern star-forming region are highlighted with an ellipse. Top right: Spectra of the eastern star-forming region (-70 pc) show strong broad emission lines with a secondary blue shifted peak most clearly visible in the [SII] emission.}
\label{fig:H210_AGN_spec2}
\end{figure}

There is also evidence that the black hole outflow is triggering the formation of star clusters in the central region of Henize 2-10. HST imaging shows that the ionized filament extends eastward from the massive black hole to a bright knot of ionized gas and site of recent star formation located 1.5" ($\sim$70 pc) away from the black hole (Figure 3). Given that our HST spectroscopy along this filament exhibits a continuous velocity pattern, which can be tracked from the black hole to the eastern star-forming region and is well described by a precessing bipolar outflow model, this strongly suggests that the outflow driven by the black hole is causally connected to the region of recent star formation. There is also a secondary, blue-shifted peak (offset by 154 km s$^{-1}$) detected in the emission lines at the location of the bright star-forming knot, suggesting the outflow is pushing the line-emitting gas clouds and influencing their kinematics (Figure \ref{fig:H210_AGN_spec2}). The double peaked lines would naturally arise as the outflow intercepted dense gas and primarily pushed it in the lateral direction rather than ahead of the flow. The handful of young star clusters associated with the star-forming knot have ages of $\sim$4 Myr (see Methods) and are predominately aligned in the north-south direction, consistent with a scenario in which they formed from gas moving in opposite directions due to the impact of the black hole outflow \cite{kharb2017double}. There is also a local peak in the gas density at the location of the star-forming knot with n$_e \geq 10^4$ cm$^{-3}$ (see Figure \ref{fig:gas_density}), consistent with the outflow compressing gas clouds and enhancing star formation in this region. It is notable that we also detect double peaked emission line profiles to the west of the massive black hole at the boundary of a dark cloud of gas and dust (region 5, see Figure \ref{fig:extract_reg}), suggesting the other side of the bipolar outflow is also intercepting dense clouds and pushing them perpendicular to the outflow direction. Just to the west of this region (region 6, Figure \ref{fig:extract_reg}), the equivalent width of the H$\alpha$ emission line indicates stellar ages $\leq$ 3 Myr, indicating a very early stage of star formation in which the infant star clusters have not had enough time to destroy or disperse the dense molecular gas from which they formed, leading to high levels of extinction \cite{beck2018dense}.  \par

The ionization conditions of the gas provide additional support for a massive black hole with an outflow. First, our HST spectroscopy at $\sim$0.1” resolution clearly reveals non-stellar ionization at the location of the central source based on the flux ratio of [OI]/H$\alpha$, which is consistent with gas photoionized by an accreting massive black hole (Figure \ref{fig:BPT}). Additionally, supernova remnants are known to be strong [SII] emitters with log([SII]/H$\alpha$) $>$ -0.521, which is not seen in the STIS spectrum of the nuclear source. The ionization conditions probed by optical diagnostics using [OIII]/H$\beta$ versus [NII]/H$\alpha$ and [SII]/H$\alpha$ do not clearly indicate a (luminous) AGN, which at first glance my seemingly rule out the presence of a massive black hole. However, there are a number of potential reasons for this apparent discrepancy.  First, [NII] and [SII] are not as sensitive to the hardness of the ionizing radiation field as [OI] and are therefore less reliable at identifying AGNs. Second, the massive black hole in Henize 2-10 is accreting at a tiny fraction of its Eddington luminosity ($\sim10^{-6}$ L$_{Edd}$) \cite{reines2011actively_Henize}, which can the impact emission lines such that they do not look like those of luminous AGNs with higher Eddington ratios \cite{trump2011accretion}. The apparent contradiction of emission line diagnostics can also arise when there is underlying star formation contributing to the spectrum. Finally, we note that there are other examples of radio-selected massive black holes in dwarf galaxies with enhanced [OI] that do not look like optical AGNs in the other diagnostic diagrams \cite{reines2020new,molina2021outflows}.  \par


We also find that the ionization conditions probed by strong emission line ratios at the location of the central source and along the filament are well described by theoretical models of shocks propagating in a high density medium \cite{allen2008mappings} (see Methods). Specifically, emission line ratios in the central $\sim$50 pc along the EW slit orientation are consistent with models of a shock with a velocity of $\sim$200 km s-1 traveling through a gas with an electron density of $\sim$1000 cm$^{-3}$ and include a ‘precursor’ of ionizing photons that travel upstream from the shock front pre-ionizing the gas. The model shock velocities that match the strong emission line ratios agree with those obtained from direct measurements of the line widths from our kinematic study described above (e.g., [OIII]), and the model gas densities are consistent with our measurements using the density sensitive line ratio [SII]6716/6731 (see Methods). Moreover, these conditions are well explained by an AGN-driven outflow mechanically exciting the interstellar medium in these regions in addition to the precursor component contributing to the photoionization of the interstellar medium. There is also evidence for shocked emission at the location of the eastern star-forming region, particularly from the secondary blue-shifted peak, in addition to photoionization from young hot stars (see Methods). Combining these results with our kinematic study indicates that the bipolar outflow generated by the central black hole is shocking the interstellar medium in the central regions of Henize 2-10, creating conditions that are favorable for positive AGN feedback \cite{silk2009global,silk2013unleashing}.  \par

AGN driven outflows have been discovered in a small sample of dwarf galaxies, though the discovery of a black hole outflow in Henize 2-10 provides the first example that is robustly spatially resolved. Moreover, Henize 2-10 differs from these other systems in several other ways. The majority of black hole driven outflows in dwarf galaxies have been found in galaxies with well-defined nuclei and optically-selected AGNs with relatively high accretion rates. These observations suggest the AGNs play a role in heating and expelling gas in the galaxies and quenching star formation, a phenomenon known as negative feedback \cite{penny2018sdss}. This is in stark contrast with Henize 2-10, which has an irregular central morphology, is intensely forming stars, and is experiencing positive feedback from a weakly accreting black hole that is luminous at radio, rather than optical, wavelengths.  \par

Indeed, “jet-mode” feedback is often associated with radio-loud AGNs accreting at low Eddington ratios, and attributed to the unbound nature of radiatively inefficient accretion flows \cite{trump2011spectropolarimetric,santoro2020agn}. Warm ionized gas outflows are observed to accompany the radio jets/outflows in some cases, particularly in young radio galaxies where nascent jets are expanding through the interstellar medium in the central regions of their hosts \cite{santoro2020agn}. In these systems, the spatial extents of the warm outflows (traced by emission lines) are similar to the radio morphologies. A similar phenomenon is observed in Henize 2-10, which is most readily seen in a comparison between the radio emission detected with the Very Large Array and Pa$\alpha$ emission detected by HST in the central few hundred pc of the galaxy \cite{reines2011actively_Henize}. However, an important difference between Henize 2-10 and powerful young radio galaxies is that the emission lines are dominated by AGN photoionization in the more massive and luminous systems, whereas the extended emission line regions in Henize 2-10 are dominated by star formation (enhanced/triggered by the black hole outflow).  Therefore, Henize 2-10 may be a low-mass, low-power analog of young radio galaxies. 

\section{Methods}

\subsection{The Controversy – Massive Black Hole or Supernova Remnant:}

\begin{figure}
\centering
\includegraphics[width=\textwidth]{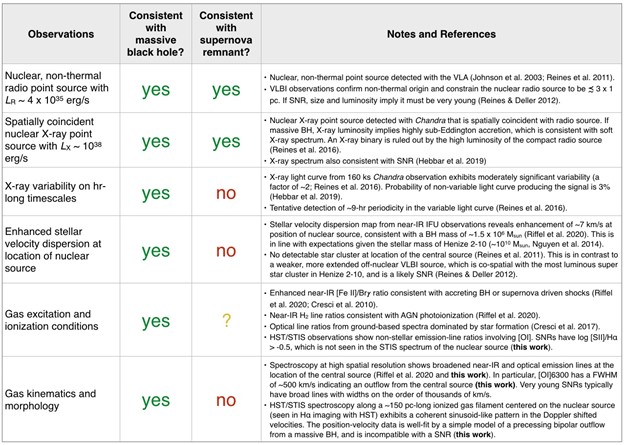}
\caption{\textbf{(Extended Data Table 1) Summary of observational results regarding the nature of the nucleus in Henize 2-10.}  A highly sub-Eddington massive black hole is consistent with all the available observations including new results presented here, while a supernova remnant is not.}
\label{tab:AGNSNR_tab}
\end{figure}

In recent years, evidence has been mounting for a massive black hole powering a low-luminosity active galactic nucleus (AGN) at the center of Henize 2-10 \cite{reines2011actively_Henize,reines2012parsec,reines2016deep,riffel2020evidence}, although a supernova remnant has been proposed as an alternative by some authors \cite{hebbar2019x,cresci2017muse}. As discussed in the main text, the radio and X-ray point source luminosities are consistent with both. A recent study \cite{hebbar2019x} argues for a supernova remnant based on their findings that the X-ray spectrum is better fit by a hot plasma model (typically used for supernova remnants) than a power-law model (typically used for luminous AGNs). However, the soft X-ray spectrum of the nuclear source in Henize 2-10 does resemble massive black holes accreting at very low Eddington fractions \cite{constantin2009probing} including Sagittarius A$^*$ in the Milky Way \cite{baganoff2003chandra}. Another study using ground-based spectroscopy favored a supernova remnant origin for the central source based on a lack of any AGN ionization signatures \cite{cresci2017muse}. However, the ground-based observations used in that work had an angular resolution of $\sim$0.7", which is not sufficient to cleanly isolate the weakly accreting black hole from nearby young ($<$5 Myr) massive (M$_* \sim 10^5 M_{\odot}$) star clusters that dominate the line ratios at this relatively course angular resolution.
There are other observational results to consider regarding the origin of the nuclear radio/X-ray source in Henize 2-10.  For example, a recent study using adaptive optics integral field spectroscopy provides evidence for gas excited by an AGN and an enhanced stellar velocity dispersion at the location of the nuclear source consistent with a $\sim10^6 M_{\odot}$ black hole, favoring the low-luminosity AGN interpretation \cite{riffel2020evidence}. There is also evidence for moderately significant variability on hour-long timescales in the X-ray light curve, which is incompatible with a supernova remnant \cite{reines2016deep,hebbar2019x}. Moreover, it is reasonable to expect that Henize 2-10 hosts a massive black hole since its overall structure resembles an early-type galaxy (albeit with a central starburst) and its stellar mass may be as high as  M$_* \sim 10^{10} M_{\odot}$ \cite{nguyen2014extended}, a regime where the black hole occupation fraction is near unity \cite{greene2020intermediate}. The central starburst complicates the identification of the weakly accreting black hole, yet a variety of multiwavelength observational results taken collectively strongly support its presence. These results are summarized in Figure \ref{tab:AGNSNR_tab} (Extended Data Table 1). A highly sub-Eddington massive black hole is consistent with all of the observations, including the new work presented here, while a supernova remnant is not.  The present study not only adds to the evidence for a massive black hole in Henize 2-10, it also demonstrates that a bipolar outflow from the black hole is enhancing/triggering star formation in its vicinity.


\subsection{STIS Observations and Data Reduction:}

Spatially resolved spectroscopic observations of the nuclear regions of the dwarf starburst galaxy Henize 2-10 were obtained using the Space Telescope Imaging Spectrograph (STIS) instrument on the Hubble Space Telescope (HST). We obtained observations with two slit orientations. The first is aligned with the quasi-linear ionized gas structure identified by Reines et al.\cite{reines2011actively_Henize} and covers the central radio/X-ray source and the bright knot of ionized gas to the east. We refer to this as the East-West (EW) orientation. The second slit orientation was placed perpendicular to the EW observation at the location of the central radio/X-ray source. We refer to this as the North-South (NS) orientation. The candidate AGN itself was too faint to acquire directly, therefore we used a target acquisition with an offset from a bright point source 7.9" to the southeast.
Spectra were taken with the G750M and G430M gratings providing medium spectral resolution (R $\sim$ 5000-6000) coverage of key optical emission lines. The central wavelengths were set at 6581 \AA \ and 4961 \AA \ for the G750M and G430M gratings, respectively. At each slit orientation, two orbits were spent in G430M and one orbit in G750M. The observations were taken with a two-point dither pattern with CR-SPLITS (multiple exposures taken to aid in cosmic ray rejection) at each position to help eliminate cosmic rays. The calibrated dithered images were combined and have a spatial resolution of $\sim$0.1", which corresponds to a physical scale of $\sim$4 pc at the distance of Henize 2-10 ($\sim$9 Mpc).


\begin{figure}
\centering
\includegraphics[width=0.9\textwidth]{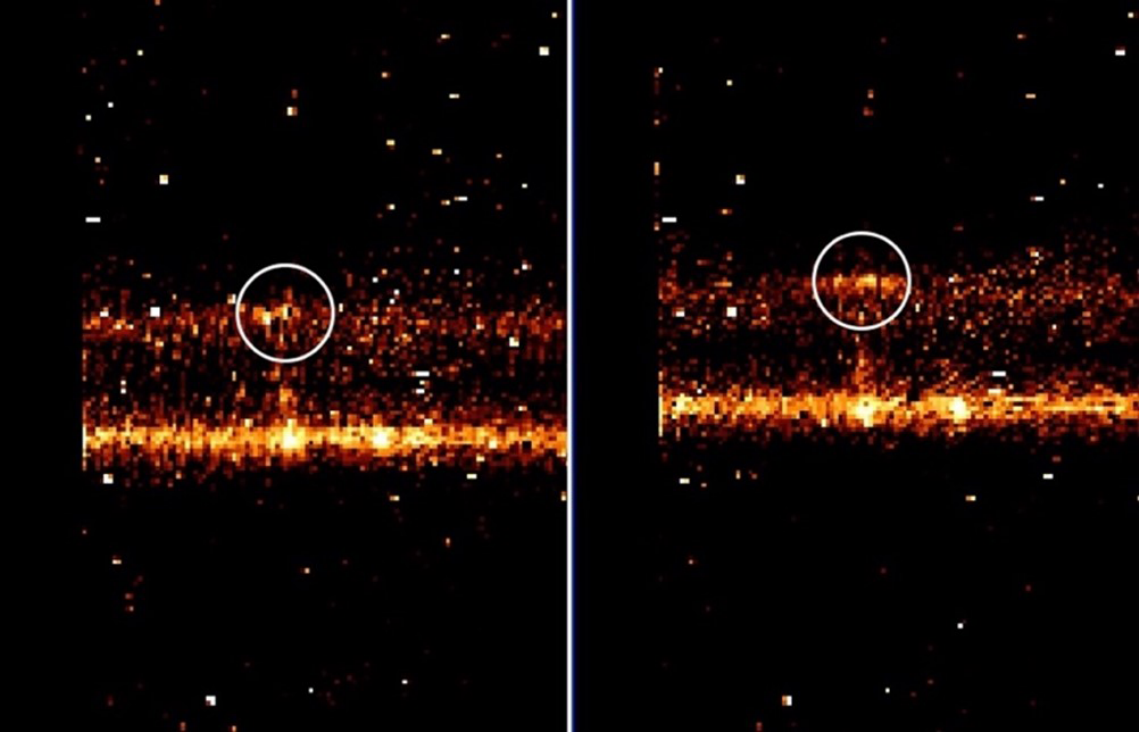}
\caption{\textbf{(Extended Data Figure 1) Raw 2D spectra showing the [OI]6300 emission line at the location of the nucleus in the EW slit orientation.} The location of the nucleus is indicated by white circles and the two images correspond to the two dithered sub-exposures.}
\label{fig:OIRAW}
\end{figure}


\subsection{Emission Line Fitting:}

\begin{figure}
\centering
\includegraphics[width=0.8\textwidth]{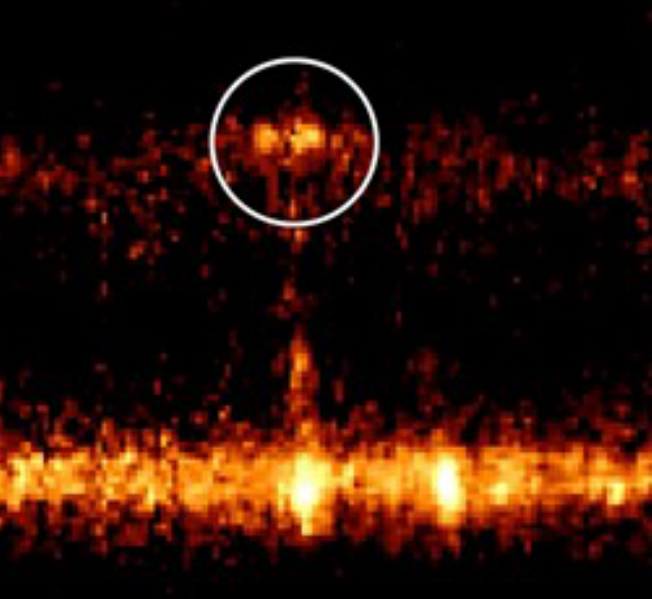}
\caption{\textbf{(Extended Data Figure 2) Combined 2D spectra showing the [OI]6300 emission line at the location of the nucleus in the EW slit orientation.} Same as Figure \ref{fig:OIRAW} but showing the reduced 2D image with the dithered sub-exposures combined. }
\label{fig:OICOMB}
\end{figure}

Before we fit emission lines in the spectra, we first modeled and removed the continuum. The reduced spectra were continuum-subtracted by masking emission line regions and fitting a low order polynomial to the continuum in each row in the spatial dimension. A low order polynomial was used given the lack of absorption features in the spectra.  We did, however, consider the potential impact of stellar absorption lines on our measurements and found that the absorption line strengths are negligible compared to the emission line strengths. Scaling a Starburst99 \cite{starburst99} model for a 4 Myr stellar population (see Stellar Ages section below) to our observed spectra, we find that the flux of H$\alpha$ absorption is smaller by a factor of 61 than the H$\alpha$ emission and the flux of the H$\beta$ absorption is smaller by a factor of 7 than the H$\beta$ emission at the location of the central source. Accounting for this absorption has a negligible effect on the line ratios of the nuclear source and does not impact the classifications based on the diagnostic diagrams.  
Once the spectra were continuum-subtracted, we fit each emission line of interest with a linear combination of Gaussian profiles to characterize the flux and estimate kinematic properties along the spatial dimension of each slit. The fitting was done using lmfit \cite{newville2016lmfit}, a non-linear least squares curve fitting package in Python. We fit each emission line with up to two Gaussian components when needed. To determine if a second Gaussian component is warranted, we require that the flux of both components be greater than the 3$\sigma$ error of the flux. This process is performed row by row in the spatial direction along each slit for emission lines of interest. During this process we fit the H$\beta$, [OIII]5007, H$\alpha$, [NII]6548/6583 and [SII]6716/6731 emission lines. We fit the H$\alpha$ and [NII] lines simultaneously, fixing the spacing between [NII] Gaussian components to their corresponding component in the H$\alpha$ emission line to laboratory values. Additionally, we tie the widths of [NII] components and fix the flux ratio of the [NII] lines to the laboratory value of 1:2.96. Similarly, the two [SII] emission lines are fit simultaneously with the spacing between Gaussian components of the two lines held fixed and the widths of the Gaussian components are tied together. 
We also fit [OI]6300 in the spectra of the nuclear source and the eastern star-forming region, but the line is too weak to be detected all along the slits. Since the [OI]6300 line has a complex profile at the location of the black hole, with possible double peaked narrow lines and a much broader component than the other emission lines, we confirmed that this was not due to an artifact in the data. In Figure \ref{fig:OIRAW}, we show close-up views of the raw 2D spectra along the EW slit position.  The two images correspond to the two dithered sub-exposures offset by 7.5 pixels. The broad [OI] line at the location of the central source is seen in both images at different positions on the detector (indicated by white circles). Note that the locations of hot pixels do not change between the two images. In Figure \ref{fig:OICOMB} we also show the final reduced 2D image with the sub-exposures combined.  The broad, double peaked nature of the [OI] line is clearly visible. We also note that [OI] is similarly broadened in the nuclear spectrum extracted from the NS slit position, although there is not an obvious double-peaked narrow line component (see Figure 2 in the main paper). In any case, broadened [OI] is clearly detected at the location of the nuclear source in both slit positions indicating an outflow on the order of ~500 km s$^{-1}$. 


\subsection{Gas Density:}

\begin{figure}
\centering
\includegraphics[width=\textwidth]{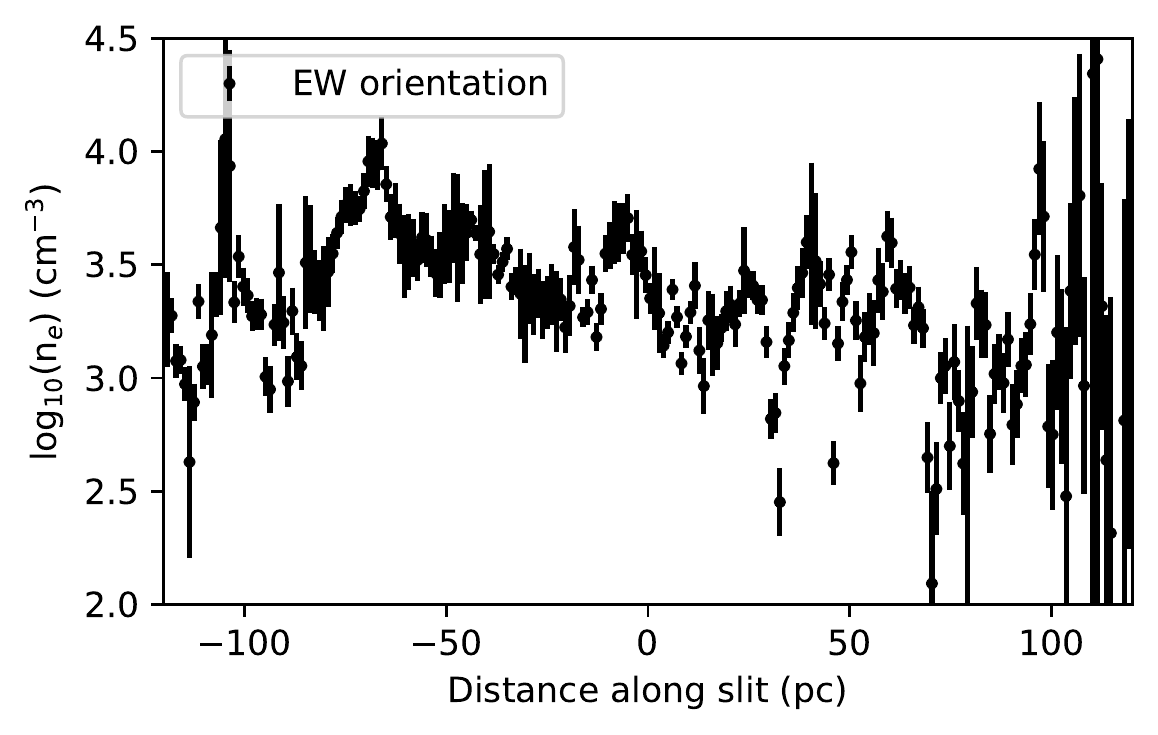}
\caption{\textbf{(Extended Data Figure 3) The electron density, n$_e$, along the EW slit orientation.} We measure the electron density along the EW slit from the ratio of [SII]6716/[SII]6731 and find the electron density ranges from $\sim10^{2.5}-10^4 cm^{-3}$ , which is within the range the [SII] ratio is sensitive to density. The high densities are consistent with those predicted by optical emission line diagnostics derived from the Allen et al. \cite{allen2008mappings} shock models. }
\label{fig:gas_density}
\end{figure}

We estimate the electron density, n$_e$, using the density sensitive line ratio [SII]6716/6731 \cite{osterbrock2006astrophysics}. This ratio is sensitive to electron densities in the range of $\sim$10$^2$-10$^4$ cm$^{-3}$. Along the EW slit orientation, we find a range of n$_e \sim 10^{2.5}-10^4 cm^{-3}$, indicating a relatively high-density gas (see Figure \ref{fig:gas_density}). These density estimates are in general agreement with the gas densities predicted by the Allen et al. \cite{allen2008mappings} shock/shock+precursor models in the central regions of Henize 2-10 as described in the next section.


\subsection{Emissin Line Diagnostics - Photoionization and Shock Models:}

\begin{figure}
\centering
\includegraphics[width=0.9\textwidth]{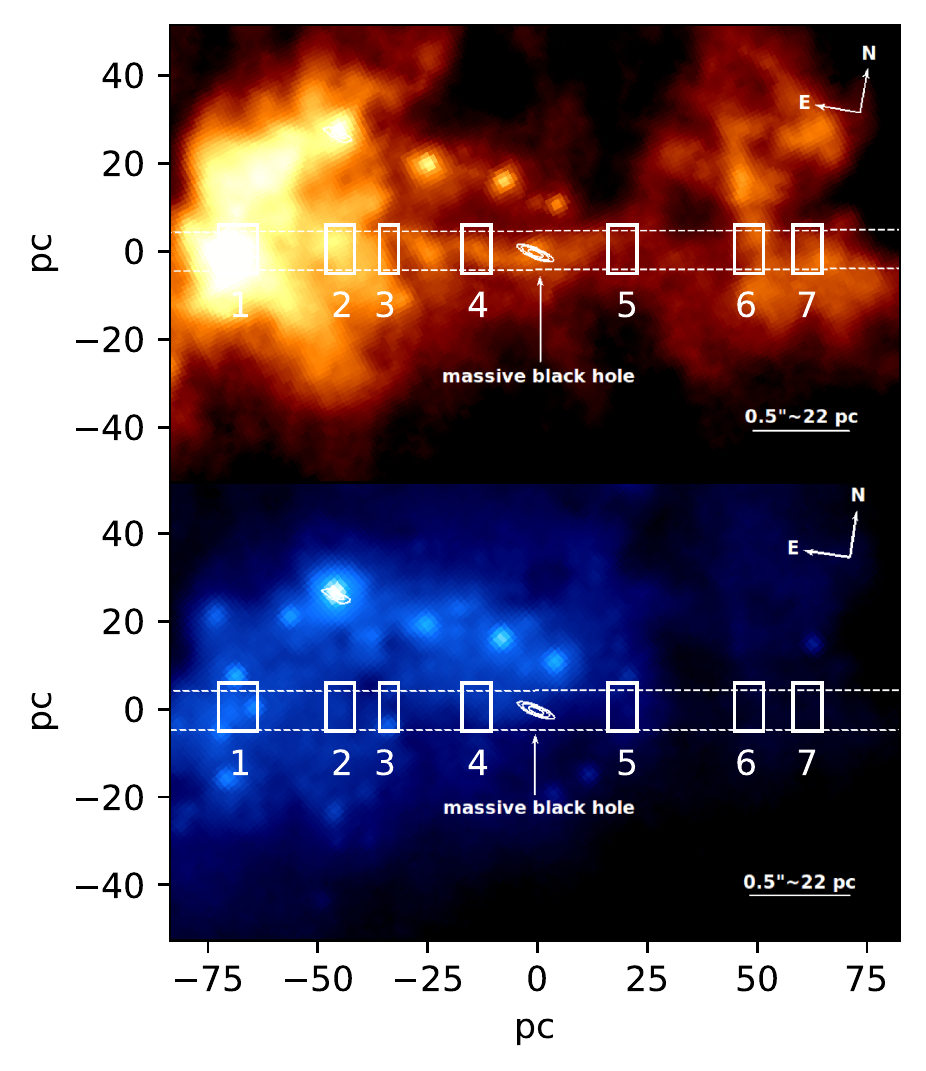}
\caption{\textbf{(Extended Data Figure 4) The spatial extraction regions taken along the EW slit orientation.} We place these regions on optical emission line diagnostic diagrams (Figures \ref{fig:BPT}-\ref{fig:ShockPre2}). Top panel: the extraction regions are shown on the narrow band H$\alpha$ + continuum image from HST to highlight the ionized gas features that several of the spatial extractions probe. Bottom panel: the extraction regions are shown on the archival 0.8 micron HST image, showing young star clusters that the EW slit orientation passes through.}
\label{fig:extract_reg}
\end{figure}

To understand the ionization mechanisms in the central regions of Henize 2-10 we compare our emission line measurements in various regions to photoionization and shock models. In addition to the central radio/X-ray source, we identified 7 regions of interest that are shown in Figure \ref{fig:extract_reg} and serve to provide a spatially resolved picture of the kinematics and ionization conditions in the central regions of Henize 2-10. The extraction regions taken along the EW slit orientation were chosen to correspond with emission features seen in the H$\alpha$ and I-band imaging from HST (young star clusters, knots of ionized gas) as well as features seen in the STIS spectroscopy (broad emission, double peaks). \par


\begin{figure}
\centering
\includegraphics[width=\textwidth]{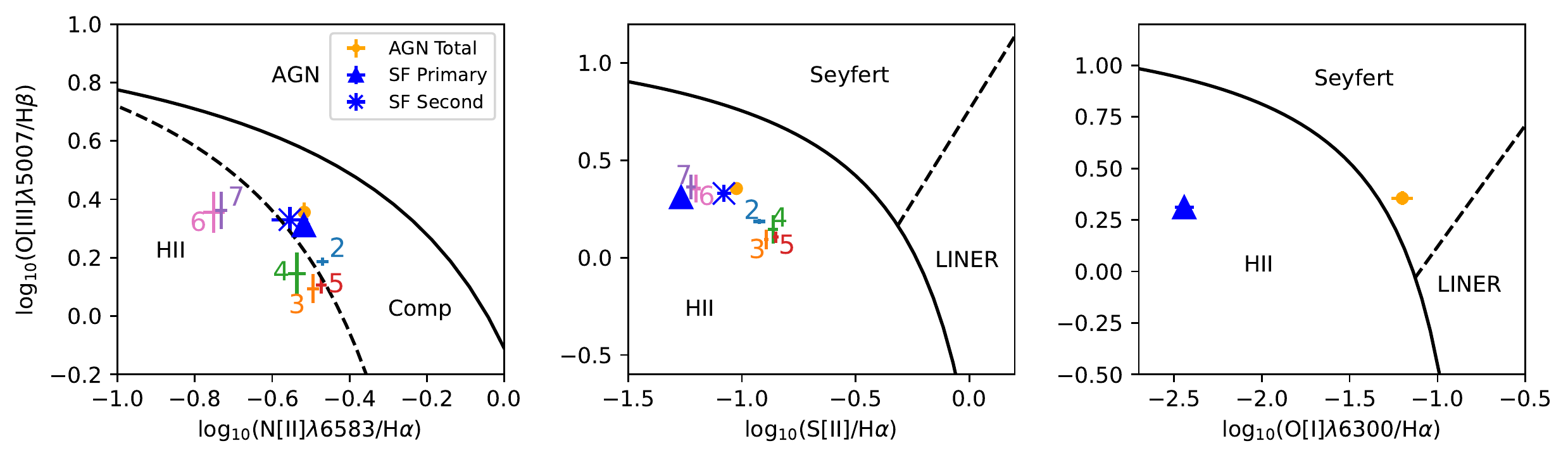}
\caption{\textbf{(Extended Data Figure 5) Narrow emission line diagnostic diagrams showing various extraction regions along the EW slit orientation (see Figure \ref{fig:extract_reg}).} The nucleus (yellow point) falls in the Seyfert region of the [OI]/H$\alpha$ diagram. The young star-forming region $\sim$70 pc to the east of the low-luminosity AGN is depicted with a blue triangle and star for the primary emission line component and the blue-shifted secondary component, respectively. [OI] is not detected in all of the regions.}
\label{fig:BPT}
\end{figure}

We first utilize the standard emission line diagnostic diagrams described by Baldwin et al. \cite{baldwin1981classification} and Veilleux et al. \cite{veilleux1987spectral} that have been expanded upon and summarized in Kewley et al. \cite{kewley2006host}. An accreting BH will produce a much harder continuum than is emitted by hot stars, and these diagrams take advantage of this fact by comparing strong emission line ratios that are close together in frequency to mitigate reddening effects. In this study we employ widely used emission line diagnostic diagrams that take [OIII]/H$\beta$ versus [NII]/H$\alpha$, [SII]/H$\alpha$, and [OI]/H$\alpha$ (see Figure \ref{fig:BPT}). In the [NII]/H$\alpha$ diagram, line-emitting galaxies separate into a V-shape \cite{kewley2006host} with star forming galaxies occupying the left most plume while AGNs occupy the right branch of galaxies. These regions are quantified by an empirical division between HII regions and emission from AGNs developed by Kauffmann et al. \cite{kauffmann2003host}. The “composite” region between this empirical division and the theoretical maximum starburst line from Kewley et al. \cite{kewley2001theoretical} indicates there is likely significant emission form both HII regions and AGNs. Like the [NII]/H$\alpha$ diagram, the [SII]/H$\alpha$ and [OI]/H$\alpha$ diagnostics provide diagnostics for differentiating between emission from HII regions and AGNs. These two diagrams add a dividing line to distinguish between emission from Seyferts and Low Ionization Nuclear Emission Regions (LINERs). LINER emission can be generated both by shocks and very hard AGN spectra and determining the primary ionization mechanism can be complicated \cite{baldwin1981classification}. It should be noted that while these diagnostic diagrams are useful for identifying regions dominated by luminous AGNs, they have limitations and can yield ambiguous results for (or completely miss) massive black holes accreting at very low Eddington ratios such as the one in Henize 2-10. Indeed, non-stellar ionization is clearly indicated in the [OI]/H$\alpha$ diagram at the location of the nuclear source, but not so for the other diagnostic diagrams (see discussion in the main paper). Figure \ref{fig:BPT} shows the [OIII]/H$\beta$ versus [NII]/H$\alpha$, [SII]/H$\alpha$, and [OI]/H$\alpha$ diagnostic diagrams for the various extraction regions along the EW slit. \par


\begin{figure}
\centering
\includegraphics[width=0.9\textwidth]{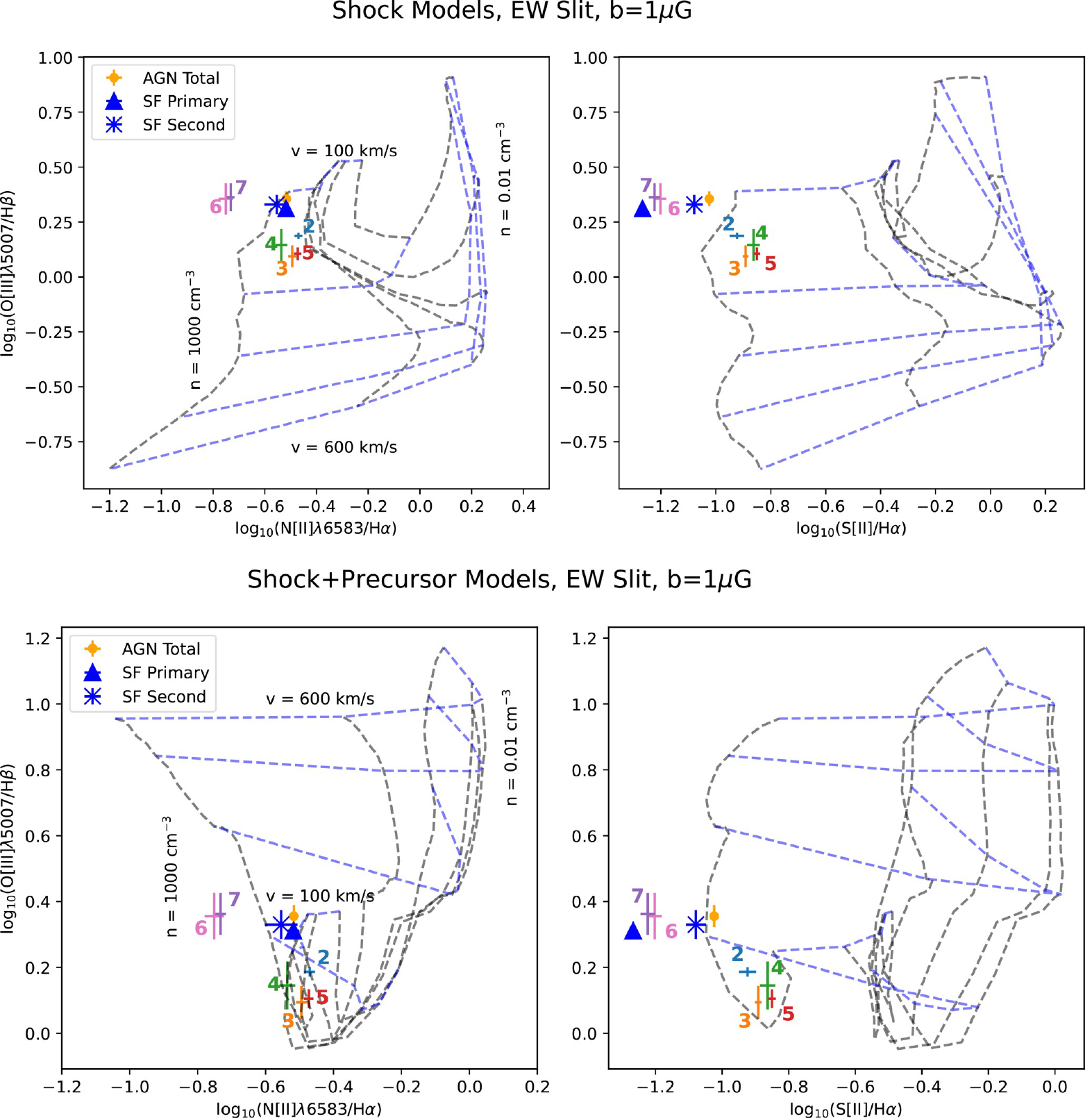}
\caption{\textbf{(Extended Data Figure 6) Optical emission line diagnostics from the shock and shock+precursor models with varying gas density.} We place the spatial extractions from the EW slit orientation shown in Figure \ref{fig:extract_reg} on a grid of shock excitation models (presented in Allen et al. \cite{allen2008mappings}) with varying gas density (n = 0.01-1000 cm$^{-3}$) and shock velocity (v = 100-600 km/s). We fix the transverse magnetic field to be b = 1$\mu$G and the assume solar metallicity.}
\label{fig:ShockPre1}
\end{figure}

\begin{figure}
\centering
\includegraphics[width=0.85\textwidth]{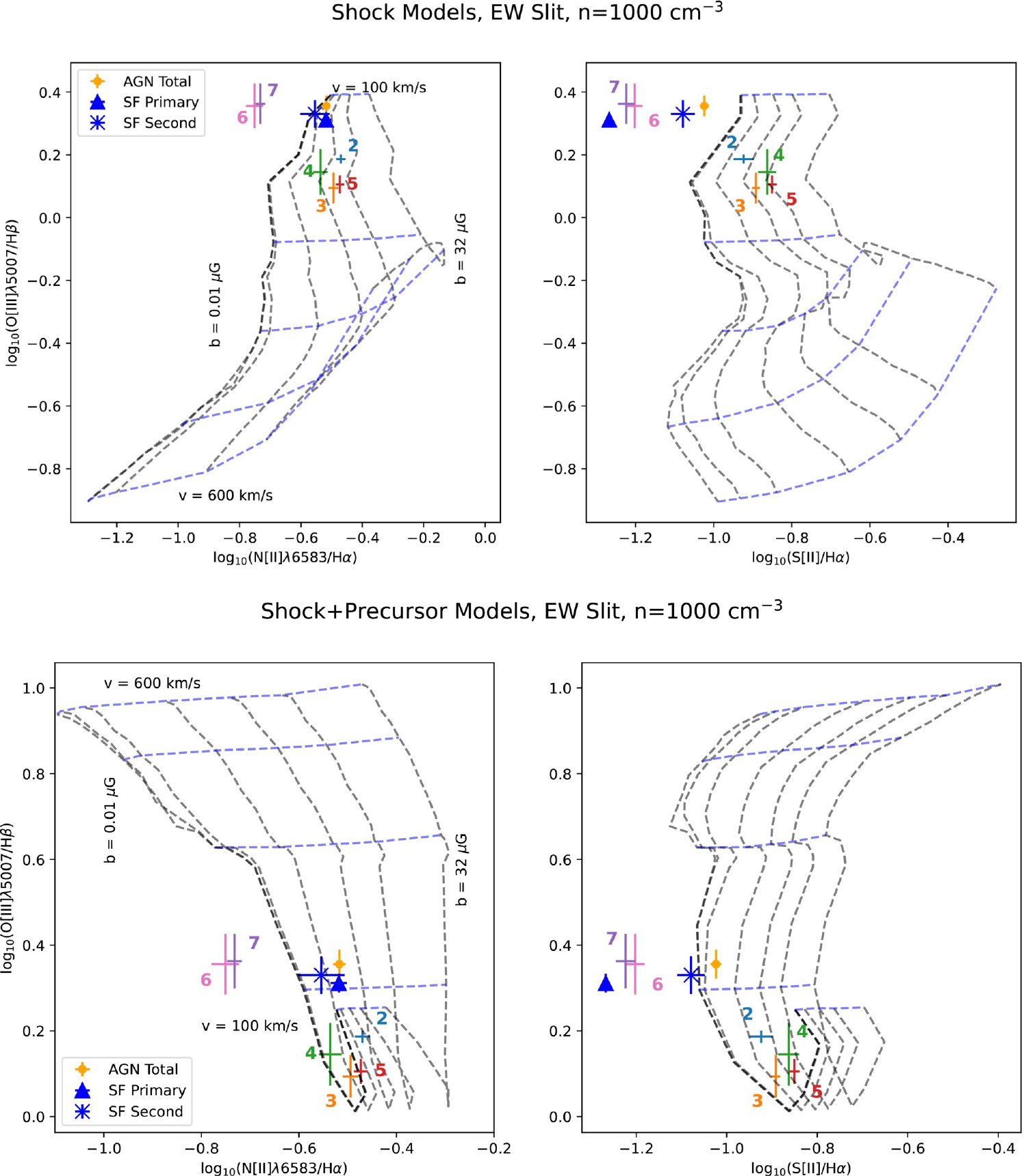}
\caption{\textbf{(Extended Data Figure 7) Optical emission line diagnostics from the shock and shock+precursor models with varying magnetic field.} The models (presented in Allen et al. \cite{allen2008mappings}) are shown as a grid with dashed blue lines indicating constant shock velocity and dashed black lines indicating constant transverse magnetic field. For these models, the density is fixed to n = 1000 cm$^{-3}$ and the transverse magnetic field parameter is allowed to vary from b = 0.01-32 $\mu$G.  }
\label{fig:ShockPre2}
\end{figure}

In addition to the diagnostic diagrams discussed above, we also investigate whether the ionization conditions seen in the central regions of Henize 2-10 can be explained by mechanical excitation from shocks. To investigate this, we employ ionization models of shock and shock+precursor emission developed by Allen et al. \cite{allen2008mappings}. These models provide emission line fluxes for ionization from a pure shock (possibly driven by an outflow from an AGN or by regions of intense star formation), where the gas is collisionally ionized, or a shock+precursor where ionizing photons produced in the shock-heated gas travel upstream and ionize the gas before the shock reaches it. We explore models with a variety of electron densities (0.01-1000 cm$^{-3}$), shock velocities (100-600 km s$^{-1}$) and transverse magnetic field strengths (0.01-32 $\mu$G). These are shown in Figures \ref{fig:ShockPre1} and \ref{fig:ShockPre2}. \par


The emission line ratios from the nuclear source are best described by the shock+precursor models with a low shock velocity (100-250 km s$^{-1}$), a high-density gas (n = 1000 cm$^{-3}$), and a low transverse magnetic field parameter (b = 0.01 – 1 $\mu$G) (Figures \ref{fig:ShockPre1} and \ref{fig:ShockPre2}). Shock+precursor models are thought to be a good description for AGN+outflow emission. Along the filament (extraction regions 2-5), the line ratios are explained well by a low velocity shock or shock+precursor model ($\sim$200 km s$^{-1}$) in a high density (n$_e \sim 1000$ cm$^{-3}$) gas with a transverse magnetic field parameter in the range of 1-10 $\mu$G. This is consistent with a scenario where the central black hole is driving a bipolar outflow that shocks the gas and dominates the ionization conditions along the filament. \par


At the region of intense star formation located $\sim$70 pc to the east of the black hole, we observe strong emission lines including a secondary, blue-shifted kinematic component. To properly fit the emission lines in this region, an additional Gaussian component is added (see Figure \ref{fig:H210_AGN_spec2} in the main paper). The separation of this component is determined using the [SII] emission line region where the secondary peak is most clearly resolved. We find that the secondary peak is offset from the primary peak by 154 km s$^{-1}$, and we use this value when fitting the other emission line regions where the secondary peak is not as well resolved. When comparing to shock(+ precursor) models, we find that the [NII]/H$\alpha$ diagram indicates a low shock velocity (100-250 km s$^{-1}$) in a high-density gas (n = 1000 cm$^{-3}$) with a low transverse magnetic field parameter (b = 0.01 – 1 $\mu$G) (see Figures \ref{fig:ShockPre1} and \ref{fig:ShockPre2}). The [SII]/H$\alpha$ diagram shows the primary emission peak is inconsistent with emission from shocks or shocks+precursor models. This indicates that the primary emission peak at this location is primarily due to star formation.  The secondary emission peak is consistent with shock+precursor models for the low velocity, high density conditions, indicating that this kinematically distinct emission component is dominated by a shock+precursor from the AGN-driven outflow.  \par

Emission from extraction regions 6 and 7 (i.e., the western star-forming region) is not consistent with any shock or shock+precursor models, which is in agreement with their location in the HII region of the Baldwin–Phillips–Terlevich (BPT) diagram. The line ratios are dominated by star formation in this region.

\subsection{Star Cluster Ages:}

We estimate the ages of the young stellar clusters that fall within the EW slit from their H$\alpha$ and H$\beta$ equivalent widths. To ascertain ages from these equivalent widths we employ simple stellar population (SSP) models from Starburst99 \cite{starburst99}. We use models from Version 7.1 with solar metallicity (appropriate for the central regions of Henize 2-10 \cite{martin2006high}), an instantaneous burst of 10$^4$ M$_\odot$ with a Kroupa IMF (0.1 – 100 M$_\odot$), the Geneva evolutionary tracks with high mass loss and the Pauldrach-Hillier atmospheres. \par

At the location of the young stellar clusters in the eastern star-forming region (region 1 in Figure \ref{fig:extract_reg}) we find an equivalent width of 478 \AA \ and 70 \AA \ for H$\alpha$ and H$\beta$ respectively. These both give stellar population age estimates of $\sim$4.3 Myr, which is in good agreement with previous estimates of the ages of other young star clusters in the region \cite{chandar2003stellar}. The ages of these clusters are larger than the crossing time for the AGN-driven outflow ($\sim$0.3 Myr), based on the minimum outflow velocity measured from emission line spectra ($\sim$200 km s$^{-1}$) and the distance between the AGN and the eastern star-forming knot ($\sim$70 pc). Therefore, the timescales allow for the AGN-driven outflow to have triggered/enhanced the formation of star clusters in the Eastern star-forming knot. \par

The EW slit orientation also passes through a young star cluster in region 3. At this location we find equivalent widths of 212 \AA \ and 41 \AA \ for H$\alpha$ and H$\beta$ respectively, both indicating an age of $\sim$5.2 Myr for the stellar cluster. Finally, region 6 in the western star-forming region has H$\alpha$ and H$\beta$ equivalent widths of 1092 \AA \ and 196  \AA, respectively. These equivalent widths indicate the stellar clusters have ages $\leq$3 Myr.

\subsection{Bipolar Outflow Model:}

\begin{figure}
\centering
\includegraphics[width=0.9\textwidth]{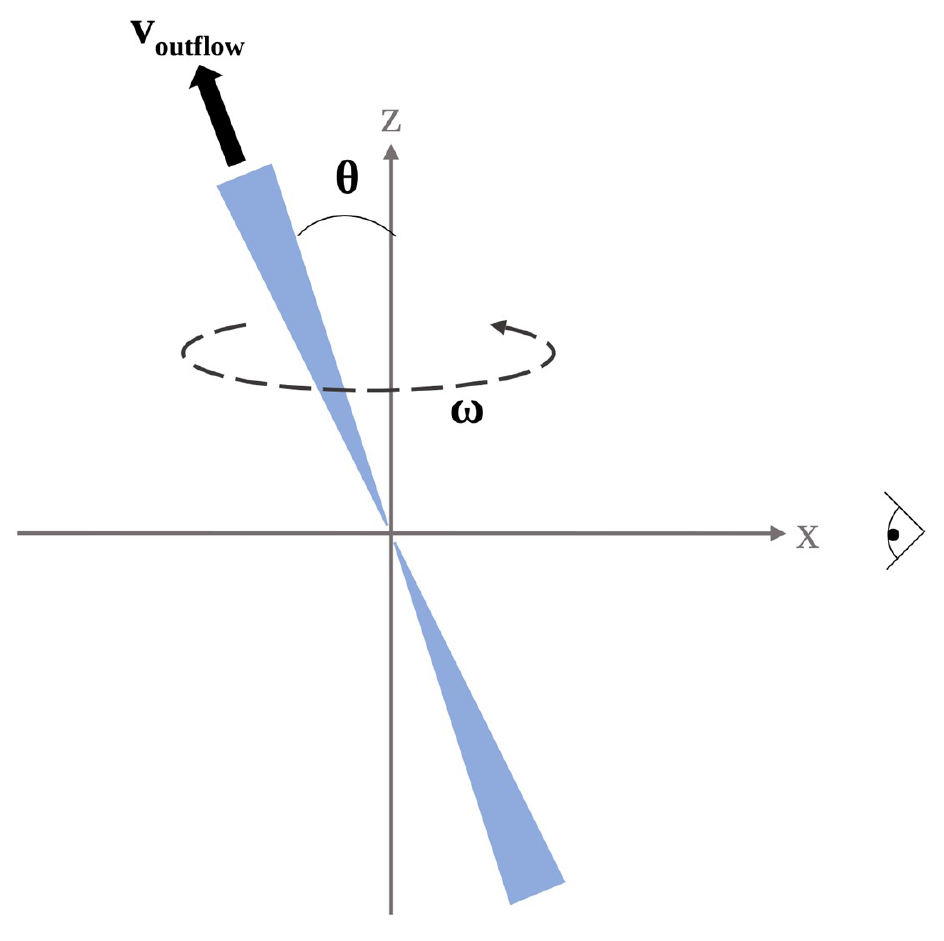}
\caption{\textbf{(Extended Data Figure 8) A diagram of the toy model of the bipolar outflow generated by the low-luminosity AGN in Henize 2-10.} Our simple model depends on the outflow velocity of the ionized gas (v$_{outflow}$), the angle the outflow makes with its precession axis ($\theta$) and the angular frequency with which the outflow precesses ($\omega$). Similar models have been used to describe the bending seen in large radio jets \cite{gower1982precessing,dunn2006precession}.}
\label{fig:outflow_model}
\end{figure}

Here we provide a derivation of the model used to describe a precessing bipolar outflow emanating from the central radio/X-ray source, which can explain the coherent velocity structure seen in the central $\sim$120 pc of the EW orientation observations. In this model we align the EW slit orientation with the $z$-axis and assume the outflow precesses about this axis with a small angle $\theta$ and an angular precession frequency $\omega$. If the gas being ejected by the outflow has velocity $v_0$, and we orient the $x$-axis to be in the direction of the observer, then the radial (Doppler shifted) velocity seen at the location of the AGN as a function of time will be given by

\begin{equation}
    v_r(t) = v_x(t) = v_0 \sin{\left(\theta\right)} \sin{\left(\omega t + \gamma\right)},
\end{equation}

\noindent
where $\gamma$ represents a phase shift that accounts for small asymmetries in the outflow profile (see Figure \ref{fig:outflow_model} for an illustration of our model).

In order to find the radial velocity as a function of distance ($z$) along the slit axis, we must consider what angle the outflow made with the (line-of-sight) $x$-axis when the gas at distance $z$ was emitted. Since this angle is time dependent as the outflow precesses, the line-of-sight velocity of the gas will depend on the orientation of the outflow at some time $t_0$ in the past. The time that has passed since the gas at distance $z$ was ejected by the outflow is determined by the $z$ velocity of the gas. Due to the symmetry of the model about the $z$-axis, the $z$ component of the gas velocity will be time independent and only depend on the angle of the outflow with the $z$-axis,

\begin{equation}
    v_z = v_0 \cos{(\theta)}.
\end{equation}

\noindent
The time, $t_0$, for gas to reach a distance $z$ along the slit is then given by

\begin{equation}
    t_0 = \frac{z}{v_0\cos(\theta)}.
\end{equation}

\noindent
We are then able to find the an expression for $v_r(z)$ by evaluating the expression for $v_r(t)$ at the time $-t_0$:

\begin{equation}
    v_r(z) = v_0 \sin{\left(\theta\right)} \sin{\left( \gamma - \frac{\omega}{v_0\cos{(\theta)}} z \right)}
\end{equation}

To fit this model to the data we require a rough estimate of the bulk outflow velocity, $v_0$. We estimate this parameter using $W80$, the velocity interval containing 80\% of the line flux, of the broad emission seen at the location of the candidate AGN.  We find $W80 \approx 200 - 500$ km s$^{-1}$ based on measurements of the [OIII]5007 and [OI]6300 lines at the location of the candidate AGN. This allows us to determine the best-fit angle of precession, $\theta$, and the frequency of precession, $f = \omega/2\pi$, to be

\begin{align}
    \theta &= 2.4^{\circ} - 6.1^{\circ} \\
    f &= 3.0 - 7.5 \ {\rm revolutions \ Myr}^{-1}
\end{align}

where the larger angle of precession and smaller frequency of precession corresponds to lower outflow velocities. We find consistent results when using the Doppler shift profile of H$\alpha$, H$\beta$ and [OIII] emission lines to fit the model derived above (the results using the [OIII] emission line is shown in Figure \ref{fig:H210_AGN_spec2}). The Doppler shift profile can be coherently traced out to 50-60 pc on either side of the candidate AGN, most definitively out to the bright eastern star forming region after which the Doppler shifts of the emission lines are influenced by the bright young stars and then shown no coherent pattern further along the slit. The coherent velocity structure seen on the scale of 100 pc is not consistent with young supernova remnant as the compact radio/x-ray source in the central regions of Henize 2-10, which provides further motivation that a low luminosity AGN is driving the outflow. 

These results are roughly consistent with jet parameters derived in other studies where precessing or reorienting jet models have been applied. The long precession period we observe ($\sim$200,000 years) is shorter by a factor of a few than those seen predicted by Dunn et al. \cite{dunn2006precession}, Nawaz et al. \cite{nawaz2016jet} and Cielo et al. \cite{cielo2018feedback} but longer by a factor of a few than those predicted by Gower et al. \cite{gower1982precessing} when jet precession is invoked to explain the complex bending and knotting seen in large radio jets.


\clearpage

\section{Acknowledgments}

We are grateful to Mallory Molina for useful discussions regarding shocks. We also thank Mark Whittle and Kelsey Johnson for their assistance with the HST/STIS proposal while AER was a graduate student at the University of Virginia, as well as subsequent discussions. Support for Program number HST-GO-12584.006-A was provided by NASA through a grant from the Space Telescope Science Institute, which is operated by the Association of Universities for Research in Astronomy, Incorporated, under NASA contract NAS5-26555. AER also acknowledges support for this work provided by NASA through EPSCoR grant number 80NSSC20M0231. ZS acknowledges support for this project from the Montana Space Grant Consortium.

\section{Author contributions statement}

ZS reduced and analyzed the STIS data and compared the results to models. AER led the HST/STIS proposal and helped with the data reduction. Both authors worked on the interpretation of the results and writing of the paper. 

\section{Data Availability Statement}
The spectroscopic data analyzed in this study are available from the Mikulski Archive for Space Telescopes (MAST), https://archive.stsci.edu/

\section{Competing Interest Statement}
The authors declare no competing interests.

\clearpage

\bibliographystyle{plain}
\bibliography{main}

\end{document}